\documentstyle[prl,twocolumn,aps,epsf,floats]{revtex}

\begin{document}

\newlength{\figurewidth}
\figurewidth=\columnwidth
\addtolength{\figurewidth}{2\columnsep}

\draft
\wideabs{
\title{Resistance behavior near the magnetic-field-tuned quantum
transition in superconducting amorphous In--O films}
\author{V.F.~Gantmakher\thanks{e-mail: gantm@issp.ac.ru},
M.V.~Golubkov, V.T.~Dolgopolov, G.E.~Tsydynzhapov, and A.A.~Shashkin}
\address{Institute of Solid State Physics, Russian Academy of
Sciences, 142432 Chernogolovka, Russia}
\maketitle

\begin{abstract}
We have studied the magnetic-field-tuned superconductivity destroying
quantum transition in amorphous In--O films with the onset of
superconductivity in zero field at about 2~K. At temperatures down to
30~mK the critical resistance $R_c\equiv R(T,B_c)$ has been found to
change approximately linearly with temperature, which is in
contradiction to a standard description where zero slope $\partial
R_c/\partial T=0$ is assumed near $T=0$. To make the data $R(T,B)$
collapse in the vicinity of transition against scaling variable
$(B-B_c)/T^{1/y}$, one has either to allow for the intrinsic
temperature dependence of $R_c$ or to postulate the critical field
$B_c$ to be temperature-dependent $\Delta B_c= B_c(T)-B_c(0) \propto
T^{1+1/y}$. We find that the state on the high-field side of the
transition can be both insulating and metallic and we determine the
critical index $y\approx 1.2$.
\end{abstract}
}

\section{Introduction}

The class of quantum phase transitions (QPT) includes a field-induced
superconductor--insulator transition (SIT) in two-dimensional (2D)
disordered super\-conduc\-tors proposed by Fisher \cite{Fisher}.
Fisher argued that the magnetic field alters a super\-conduc\-ting
state of a disordered film at low fields, through a metallic one with
the universal sheet resistance $R_{\rm un}$ close to $h/4e^2\simeq
6.4\,$k$\Omega$ at the critical field $B=B_c$, to an insulating state
at fields $B>B_c$. The SIT is supposed to be continuous with the
correlation length $\xi$ diverging as $\xi=\xi_0(B-B_c)^{-\nu}$,
where the critical index $\nu>1$. It was argued theoretically
\cite{Girvin} and confirmed experimentally for amorphous films
\cite{HebPaa,Kapit} that the dynamical critical index $z$, which
determines the characteristic energy $U\sim\xi^{-z}$, is equal to
$z=1$. At finite temperatures, when plotted against scaling variable
$u$,

\begin{equation}      \label{x=}
u=(B-B_c)/T^{1/y}, \qquad y=z\nu,
\end{equation}
all data $R(T,B)$ near the transition point ($0,B_c$) should fall on
a universal curve \cite{Fisher}

\begin{equation}      \label{f=}
R(T,B)\equiv R_cr(u),\qquad r(0)=1.
\end{equation}

Experimental studies \cite{HebPaa,Kapit} demonstrated the collapse of
$R(T,B)$ as a function of $u$ and apparently supported this model.
Still, these raise some questions. In the high-magnetic-field limit,
amorphous Mo--Ge films from Ref.~\cite{Kapit} displayed only 5\%
increase of the resistance with ten-fold decrease of temperature and
behaved like a metal with small quantum corrections to the
resistance. On In--O films the quasireentrant scenario was observed
\cite{HebPaa}: at fields below $B_c$, the derivative $\partial
R/\partial T$ was positive near the superconducting transition
temperature $T_c$ but became negative at lower temperatures (see
below, next section).

Here, we perform the detailed study on amorphous In$_2$O$_x$ ($x<3$)
films looking for experimental evidence as concerns unclear aspects
of the Fisher transition. The main question is whether the
field-induced phase transition in amorphous In$_2$O$_x$ films is
indeed a continuous QPT at zero temperature \cite{QPT} or a broadened
thermodynamic transition with temperature-dependent critical field
$B_{c2}(T)$ as known from the theory of superconductivity. A
principal argument in favour of QPT would be the scaling relations,
i.e., $R(T,B)$ should take the form (\ref{f=}).

The scaling relations have two constituents. First, the derivative on
the isomagnetic curve $R(T,B_c)$ at $T=0$ should equal zero
\cite{Fisher}:

\begin{equation}			\label{partial}
R(T,B_c)\equiv R_c(T)=R_c\left(1+O(T^2)\right), \quad\
\frac{\partial R}{\partial T}(0,B_c)=0.
\end{equation}
This implies that isotherms $R(B)\Bigl|_{T={\rm const}}$ intersect at
the same point $(B_c,R_c)$ for all $T$ in the scaling region.
Expanding the isotherms $R(B)$ in the vicinity of $B_c$

\begin{equation}		\label{deriv}
R(B)=R_c+\left(\frac{\partial R}{\partial B}\right)_{B_c}(B-B_c)
\end{equation}
we get from Eq.~(\ref{f=}) the second component of the scaling
relations

\begin{equation}		\label{sca}
\left(\frac{\partial R}{\partial B}\right)_{B_c}\propto T^{-1/y}.
\end{equation}
From Eqs.~(\ref{deriv},\ref{sca}) it follows that near the transition
point $(0,B_c)$ the universal curve is a linear function

\begin{equation}
r(u)=1+\beta u .\label{uni}\end{equation}
The procedure of experimental tests is defined by
Eqs.~(\ref{partial}) and (\ref{sca}). One has to find either the
isomagnetic curve with zero slope near $T=0$ or the crossing point of
low-temperature isotherms to derive $B_c$ and $R_c$. The value of $y$
is determined from the temperature dependence of the derivative
(\ref{sca}). The last step is to determine the scaling region from
plot (\ref{f=}).

We intend to use this procedure for samples that are farther from the
zero-field SIT as compared to Ref.~\cite{HebPaa}. In amorphous
In$_2$O$_x$ films the carrier density $n$ can be decreased
(increased) by increasing (decreasing) the oxygen content; the
process is reversible provided the material remains amorphous
(experimental details are discussed in the next section, together
with the role of granularity). As the density $n$ increases, the
sample shifts deeper into the superconducting region. This leads to
increasing the conductance of the high-field state, which is expected
in theory \cite{Fisher} to be insulating, and the superconducting
transition temperature $T_{c0}$ at zero field, and the critical
magnetic field $B_c$. As a result, one may expect to find the
fingerprints of an ordinary thermodynamic transition that certainly
exists somewhere very deeply in this region. It is shown in section
\ref{res} that shifting inward the superconducting region gives rise
to the appearance of a linear term in Eq.~(\ref{partial}) so that the
scaling representation (\ref{f=}) fails. Nevertheless, this can be
restored in generalized form if we compensate for the linear term in
$R$ by introducing the temperature dependence of either $B_c$ in
Eq.~(\ref{x=}) or $R_c$ in Eq.~(\ref{f=}). In this way we confirm
that the field-tuned quantum SIT does occur in our films.

In section \ref{IV} we analyze whether the observed QPT is the one
described by Fisher \cite{Fisher} or it belongs to a more general
group. We demonstrate that (i) the critical resistivity $R_c$ depends
on carrier density; (ii) the high-field state can be metallic as well
as insulating; and (iii) the reduced dimension is not of crucial
importance. In particular, the film magnetoresistance is not very
sensitive to the magnetic field direction. The obtained data on
zero-temperature QPTs are summarized to form a phase diagram. We
conclude that for field-tuned quantum SIT a strong disorder is
important.

\section{Experimental}
\subsection{Samples and measurements}

The experiments were performed on amorphous In--O films, 200\,\AA\
thick \cite{InO}. Such a material proved to be very useful for
investigations of the transport properties near the SIT
\cite{HebPaa,ShOv,HP,Kim,KimLee,gg1,gg2}. Oxygen deficiency compared
to fully stoichiometric insulating compound In$_2$O$_3$ causes the
film conductivity. By changing the oxygen content one can cover the
range from a metallic superconducting material to an insulator with
activated conductance \cite{ShOv}. The procedures to change the film
state are described in detail in Refs.~\cite{gg1,gg2}. To reinforce
the superconducting properties of our films and to increase $T_{c0}$
we used heating in vacuum up to a temperature from the interval 70 --
110$^0$C until the sample resistance saturated or even longer. To
shift the sample properties in the opposite direction we made
exposure to air at room temperature.

The low-temperature measurements were carried out by a four-terminal
lock-in technique at a frequency of 10~Hz using two experimental
setups: a He$^3$-cryostat down to 0.35\,K and Oxford TLM-400 dilution
refrigerator in the temperature interval 1.2\,K -- 30\,mK. The sample
resistance $R(T,B)$ was studied as a function of $T$ at fixed $B$ in
the He$^3$-cryostat and as a function of $B$ at fixed $T$ in the
dilution refrigerator. The ac current was equal to 1~nA and
corresponded to the linear regime of response. The aspect ratio of
the samples was close to one. Transferring the sample from the
He$^3$-cryostat into the dilution refrigerator or back required for
it to stay at room temperature for half an hour, at least. Since the
sample absorbed oxygen, to restore its original state we had to
repeat the heat treatment before cooling down. Although significant,
the experiments in He$^3$-cryostat were preliminary. After describing
them we will concentrate mainly on the results obtained at the lowest
temperatures.

\subsection{Quasireentrant and ideal transitions}

\begin{figure}[tb]
\vbox{
\epsfxsize=\figurewidth
\centerline{\epsfbox{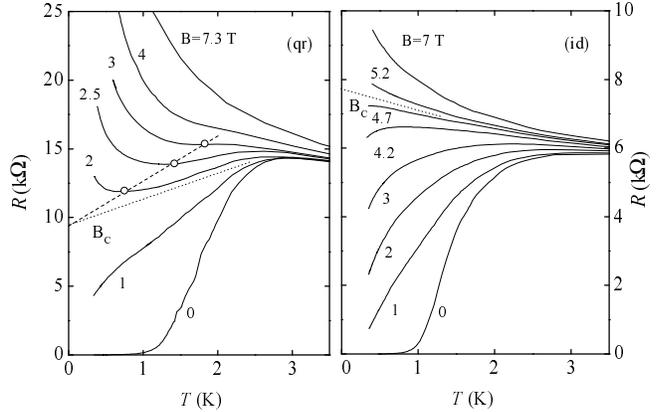}}
\caption{Temperature dependences of the film resistance in various
magnetic fields for quasireentrant (qr) and ideal (id) states. The
dashed line represents a linear extrapolation of the resistance
minima (circles) to $T=0$. The separatrices $R(T,B_c)$ are shown by
dotted lines.}
\label{11}
}
\end{figure}

One can distinguish two kinds of sets of 2D field-tuned SIT curves
$R(T)$ by their behavior below the transition temperature $T_{c0}$
(Fig.\,\ref{11}). The first type is called quasireentrant transition,
see the left part of Fig.\,\ref{11} labelled (qr); samples with such
a behavior were used in Refs.~\cite{HebPaa,Kim}. In weak fields
$R(T)$ is a maximum near $T_{c0}$ and decreases with lowering $T$
until it reaches zero or a finite value, whereas in strong fields the
derivative $\partial R/\partial T$ is negative everywhere. Over the
intermediate field range each of the curves $R(T)$ shows a maximum at
$T_{\rm max}\approx T_{c0}$ and a minimum at a temperature $T_{\rm
min}<T_{\rm max}$ that becomes lower as the field is decreased. The
second type is referred to as ideal transition (the right part of
Fig.\,\ref{11} labelled (id)). There are no curves with minima at
all; as the field is increased, the $R(T)$ maximum shifts to lower
$T$ until it disappears.

According to Ref.~\cite{Gold}, the low-temperature minimum for the
quasi\-reentrant transition is caused by inhomogeneities and
single-particle tunneling between superconducting grains. This
statement is supported by the fact that several additional hours of
annealing the sample in vacuum after its resistance has already
saturated favour the ideal behavior. We note that amorphous In--O
films of Ref.~\cite{HebPaa} revealed the quasireentrant transition
and amorphous Mo--Ge films of Ref.~\cite{Kapit} demonstrated the
ideal transition.

\begin{table}
\vbox{
\caption{Parameters of ideal states of the sample. $R_r$ is the
resistance at room temperature, the definition of $R_c$ and $B_c$ is
described in section \protect\ref{IIIa}, $\alpha$ is the slope of the
curve $R(T,B_c)/R(0,B_c)$ at $T=0$.}
\begin{tabular}{l|cccc}
State&$R_r$, k$\Omega$&$R_c$, k$\Omega$&$B_c$, T&$\alpha$, K$^{-1}$\\
\tableline
1&3.4&7&2&$\approx 0$\\
(id)&3.1&8.2&5.3&0.16\\
2&3.0&9.2&7.2&0.8\\
3\tablenote{Magnetic field was parallel to the
film.}&3.2&10&7.9&0.32\\
\end{tabular}
\label{t1}
}
\end{table}

Below we describe the results obtained on four states of the same
film with ideal transitions. The parameters of these states are
listed in Table~\ref{t1}. The state labelled (id) was studied in
He$^3$-cryostat, the rest three were obtained by heat treatment after
mounting into the top loading system of dilution refrigerator. The
metallic properties of the state can be characterized by its room
temperature resistance $R_r$. Assuming that disorder for all states
is approximately the same, we have for the carrier density $n\propto
1/R_r$, i.e., the smaller $R_r$, the stronger the superconducting
properties and, hence, the larger the value of $B_c$.

\section{results}
\label{res}
\subsection{Separatrix in the ($T,R$) plane; measurements above
0.3\,K}

Our procedure for determining the values of $B_c$ and $R_c$ is
schematically illustrated for the quasireentrant transition in
Fig.\,\ref{11}(qr). A linear extrapolation of the minimum positions
$R_{\rm min}(T)$ to $T=0$ yields the limiting $R_c$ value as shown by
dashed line in the figure. The expected separatrix $R(T,B_c)$
indicated by dotted line includes a linear term $R_c\alpha T$, which
is in contradiction to the relation (\ref{partial}).

\begin{figure}[tb]
\vbox{
\epsfxsize=\figurewidth
\centerline{\epsfbox{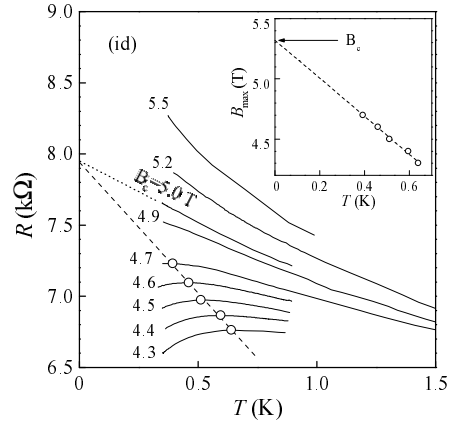}}
\caption{Blow-up of the central part of Fig.~\protect\ref{11}(id)
presented with smaller field step. The linear extrapolations to $T=0$
of the resistance maxima (circles) and of the isomagnetic curve at
$B=5.0\,$T are indicated by dashed and dotted lines, respectively.
Inset: linear extrapolation of the maximum field $B_{\rm max}$ to
$T=0$.}
\label{12}
}
\end{figure}

A similar procedure is applied to the fan of curves for the ideal
transition, see Figs.\,\ref{11}(id) and \ref{12}. The limiting $R_c$
value is obtained by extrapolating linearly to $T=0$ the maximum
positions $R_{\rm max}(T)$ (dashed line in Fig.\,\ref{12}). The
dependence $R(T)$ at $B=5\,$T in Fig.\,\ref{12} is close to a
straight line that separates curves with positive and negative second
derivatives. Its extension to zero temperature (dotted line) comes
practically to the same point $R_c$ and its slope is finite, although
of opposite sign compared to Fig.\,\ref{11}(qr). Since the maximum
position in the ($T,R$) plane is determined by magnetic field,
another way to find $B_c$ is to use a similar extrapolation to $T=0$
of the temperature dependence of the maximum field $B_{\rm max}(T)$
(inset to Fig.\,\ref{12}). A bit higher value of $B_c=5.3$\,T gives
estimate of the uncertainty of the extrapolation from above 0.3~K.

From an experimental point of view, it seems natural for the curve
$R(T,B_c)$ to have a finite derivative at $T=0$ (dotted lines in
Fig.\,\ref{11}), which implies that the scaling relations
(\ref{partial}) -- (\ref{sca}) fail. This as well as the 6\%
discrepancy between the $B_c$ values (see Fig.~\ref{12}) would
possibly mean that the temperatures used are too high. Below we
present the results of similar experiments in a dilution
refrigerator.

\subsection{Separatrix in the ($T,R$) plane; measurements below
0.3\,K}
\label{IIIa}

The sample mounted in the top loading system of dilution refrigerator
was brought by heating first into state 1 with relatively low
transition temperature $T_{c0}$ and critical field $B_c=2.2$\,T
(Table~\ref{t1}). State 2, which is deeper in the superconducting
region, was attained by means of additional heat treatment. In this
section we will consider largely the second state of the sample.

\begin{figure}[tb]
\vbox{
\epsfxsize=\figurewidth
\centerline{\epsfbox{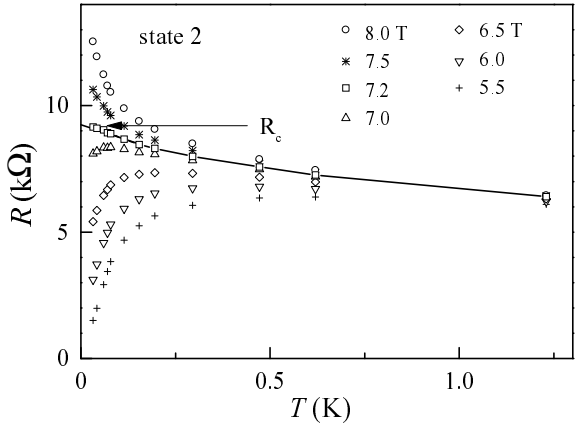}}
\caption{Temperature dependence of the state 2 resistance at
different magnetic fields. The separatrix $R(T,B_c)$ (solid line) has
nonzero slope near $T=0$.}
\label{13}
}
\end{figure}

\begin{figure}[tb]
\vbox{
\epsfxsize=\figurewidth
\centerline{\epsfbox{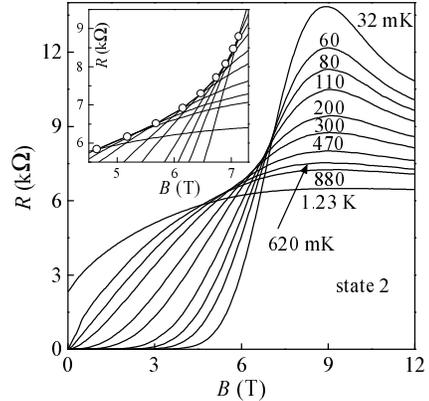}}
\caption{Isotherms in the ($B,R$) plane for state 2. The curve
intersection region is blown up in the inset. The circles mark the
crossing points of isotherms with neighboring temperatures.}
\label{14}
}
\end{figure}

Fig.\,\ref{13} displays the fan of isomagnetic curves for state 2.
As seen from the figure, maxima on the curves $R(T)$ can be traced
down to 40\,mK so that the above procedure for determining $R_c$ and
$B_c$ is applicable. One can see from Fig.\,\ref{14} that isotherms
of state 2 do not cross at the same point but form a line.
Apparently, if two isotherms $R(B)\Bigr|_{T_1}$ and
$R(B)\Bigr|_{T_2}$ intersect at a point ($B_i,R_i$), the isomagnetic
curve $R(T)\Bigr|_{B_i}$ reaches its maximum at $T\approx
(T_1+T_2)/2$.

\begin{figure}[tb]
\vbox{
\epsfxsize=\figurewidth
\centerline{\epsfbox{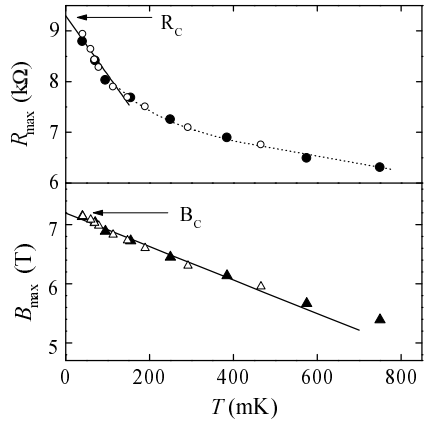}}
\caption{Change of the maximum resistance and field with temperature
as determined from the data in Fig.~\protect\ref{13} (open symbols)
and Fig.~\protect\ref{14} (filled symbols). The values of $R_c$ and
$B_c$ are obtained with the help of linear extrapolations (solid
lines). The dotted line is a guide to the eye.}
\label{15}
}
\end{figure}

The functions $R_{\rm max}(T)$ and $B_{\rm max}(T)$ are depicted in
Fig.\,\ref{15}. The open symbols correspond to the maximum positions
on isomagnetic curves (Fig.\,\ref{13}) and the filled symbols
represent the data obtained from curve intersections in
Fig.\,\ref{14}. Although neither of these functions tends to saturate
at low $T$, there is a trend for them to reach finite values $R_c$
and $B_c$ at $T\rightarrow 0$ (Fig.\,\ref{15}). The point
($T,R$)=($0,R_c$) is an unstable fixed point that belongs to the
separatrix $R(T,B_c)$ with nonzero slope near $T=0$, see
Fig.\,\ref{13}. Below we generalize the scaling relations
(\ref{partial}) -- (\ref{sca}) to make allowance for the observed
linear term.

\subsection{Generalized scaling relations}

\begin{figure}[tb]
\vbox{
\epsfxsize=\figurewidth
\centerline{\epsfbox{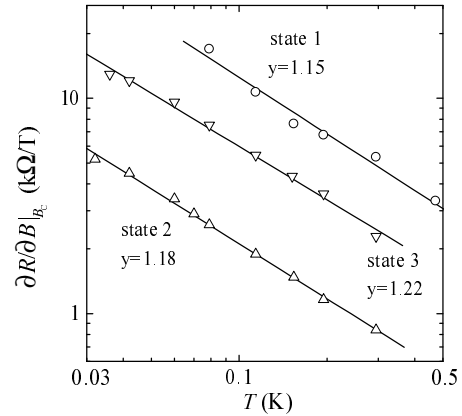}}
\caption{Behavior of $\partial R/\partial B\Bigr|_{B_c}$ with
changing temperature for states 1, 2, 3. The values of power $y$ are
indicated.}
\label{scale_y}
}
\end{figure}

Having determined the state 2 parameter $B_c=7.2$\,T we obtain the
power $y\approx 1.2$ from logarithmic plot of $\partial R/\partial
B\Bigr|_{B_c}$ vs $T$ (Fig.\,\ref{scale_y}). This critical index is
in agreement with data on the SIT $y\approx 1.3$ of
Refs.~\cite{HebPaa,Kapit} as well as on the metal-insulator
transition in 2D electron systems in semiconductors $y\approx 1$
\cite{scal,scal1} and $y\approx 1.6$ \cite{Pud}.

\begin{figure}[tb]
\vbox{
\epsfxsize=\figurewidth
\centerline{\epsfbox{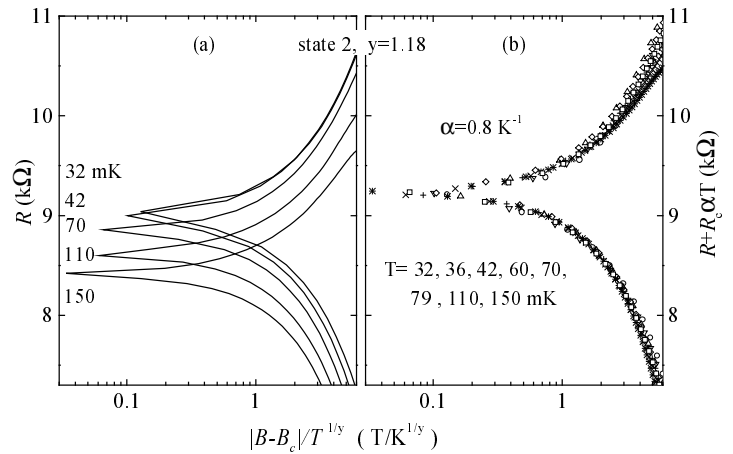}}
\caption{Scaling dependences of $R$ (a) and $\tilde R$ (b) for state
2.}
\label{16}
}
\end{figure}

The finite slope of the separatrix $R(T,B_c)$ near $T=0$

\begin{equation}			\label{gener}
R(T,B_c)\equiv R_c(T)=R_c\left(1-\alpha T+O(T^2)\right)
\end{equation}
certainly prevents the data from collapsing onto one curve when
plotted against scaling variable (\ref{x=}), see Fig.\,\ref{16}(a).
To restore the scaling behavior the linear term in Eq.~(\ref{gener})
has to be compensated. This can be done by introducing the
temperature dependence of either $R_c$ or $B_c$.

\begin{figure}[tb]
\vbox{
\epsfxsize=\figurewidth
\centerline{\epsfbox{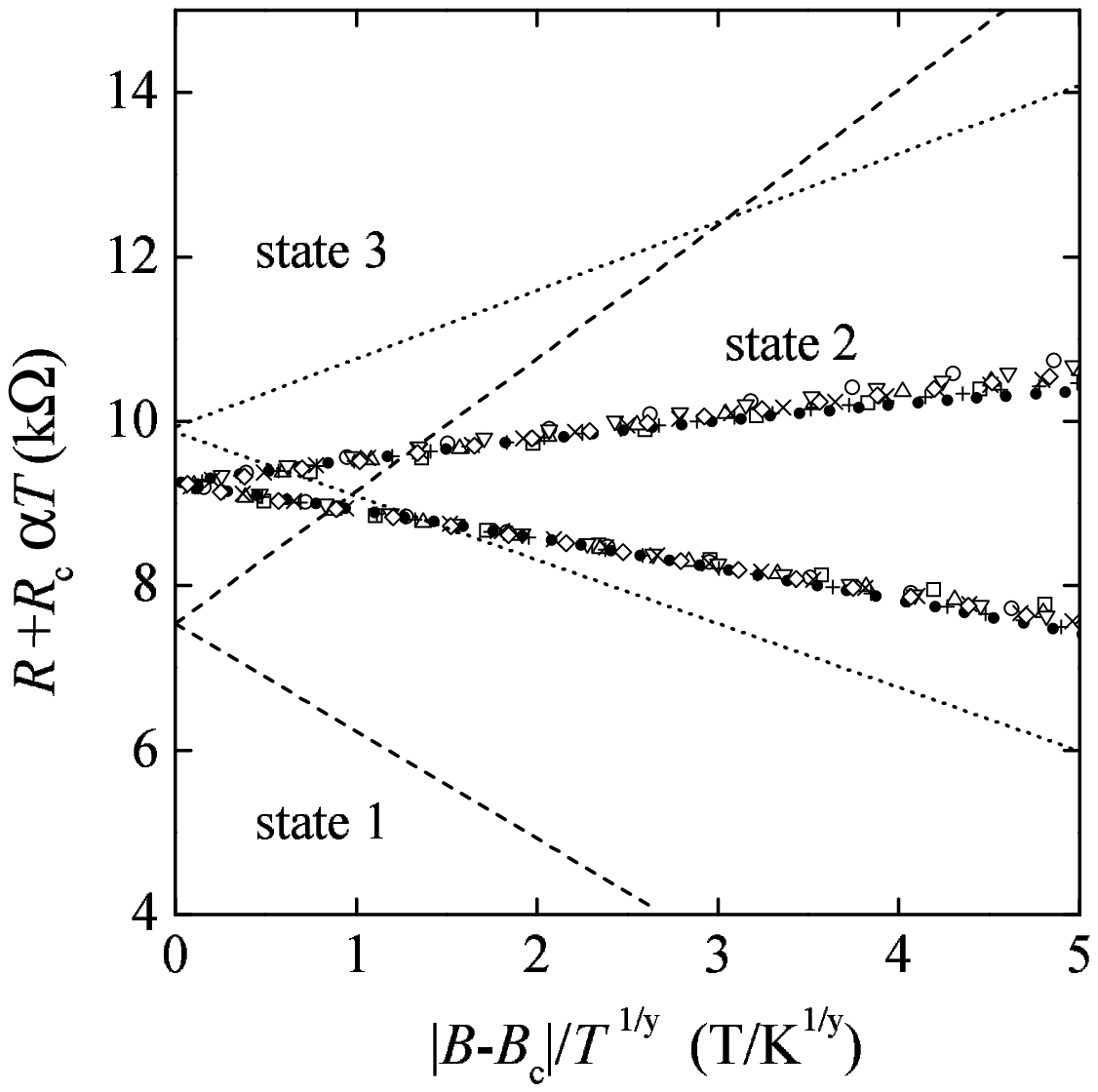}}
\caption{The same data as in Fig.~\protect\ref{16}(b) plotted on a
linear scale. Also shown are the averaged data for states 1 (dashed
lines) and 3 (dotted lines).}
\label{lin_sc}
}
\end{figure}

Assuming that the finite separatrix slope originates from the
intrinsic dependence of $R_c$ on temperature, we add the linear term
$R_c\alpha T$ ($|\alpha T|\ll 1$) to $R(T,B)$

\begin{equation}				\label{Rtilde}
\tilde R=R(T,B)+R_c\alpha T, \qquad
\partial\tilde R/\partial B\Bigr|_{B_c}\equiv\:\partial R/\partial
B\Bigr|_{B_c}
\end{equation}
and test the scaling form of $\tilde R$. By varying the coefficient
$R_c\alpha$ we reach the optimum collapse of the data onto a single
curve to determine $R_c=9.2$~k$\Omega$, see Fig.\,\ref{16}(b). This
$R_c$ value agrees well with the one obtained from a linear fit of
the data in Fig.~\ref{15} over the same temperature interval 30 to
150~mK. That the resistances $R$ and $\tilde R$ are traditionally
plotted in Fig.\,\ref{16} as a function of $\log(|u|)$ masks the fact
that the universal function $r(u)$ is merely a linear function
(Fig.\,\ref{lin_sc}) as expected from Eq.~(\ref{uni}).

\begin{figure}[tb]
\vbox{
\epsfxsize=\figurewidth
\centerline{\epsfbox{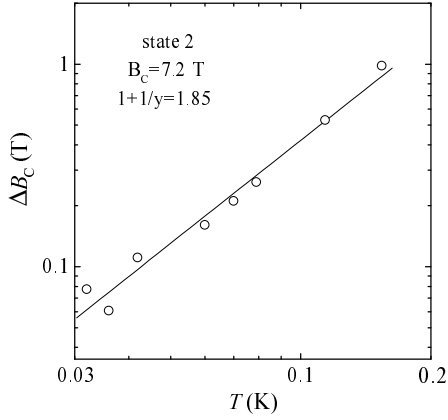}}
\caption{Change of $\Delta B_c$ defined by Eq.~(\protect\ref{B_c})
with temperature for state 2. The slope of the straight line is
indicated.}
\label{dBc}
}
\end{figure}

Another way to make up for the linear term in Eq.~(\ref{gener}) is to
introduce in Eq.~(\ref{x=}) the temperature dependent field $B_c(T)$

\begin{equation}				\label{B_c}
\Delta B_c=B_c(T)-B_c(0)\propto T^{1+1/y}
\end{equation}
defined through the constancy of $R_c$

\begin{equation}
R\left(T,B_c(T)\right)=R_c={\rm const}, \label{const}\end{equation}
see Fig.\,\ref{dBc}. In contrast to the normal behavior of the
critical fields in superconductors, so-defined field $B_c$ increases
with temperature.

The above ways to compensate for the linear term correspond to shifts
of isotherms either along the $R$-axis or along the $B$-axis (see
Fig.~\ref{14}) such that in the vicinity of transition a common
crossing point is attained. While formally these are equivalent, we
cannot tell which manner is correct from a theoretical viewpoint and
we address this problem to theory.

\section{Discussion}
\label{IV}

In the preceding section we claim that the SIT observed in our
amorphous In--O films is a quantum transition because it obeys the
generalized scaling relations. Below we shall discuss whether it
meets restrictions imposed by theory \cite{Fisher} or its character
is more general. The following items are highlighted: (i)
universality of the critical resistance value at the transition; (ii)
insulating features of the high-field phase; and (iii) parameters
that determine the 2D character of our films. The phase diagram of
the transition and the role of disorder are discussed.

\subsection{The critical resistance value}

A theoretical prediction \cite{Fisher} for the universal value of
critical resistance $R_{un}\approx 6.4$~k$\Omega$ was not confirmed
experimentally so far (see, e.g., Ref.~\cite{Kapit}). In our
experiment, where possible influence of geometrical factors was
excluded, the values of $R_c$ are found to be different for different
states of the sample, see Table~\ref{t1} and Fig.\,\ref{lin_sc}.
Moreover, comparison of states 1, (id), and 2 of our film reveals the
correlation between the values of $R_r$, $R_c$, and $B_c$. As
mentioned above, the resistance $R_r$ is supposed to be inversely
proportional to the carrier density $n$. The larger $n$, the deeper
the state in the superconducting region, leading to the increase of
both $B_c$ and $R_c$. The parameters for state 3 obtained in field
parallel to the film do not follow this tendency.

One can see from Fig.\,\ref{lin_sc} that the coefficient $\beta$ in
Eq.~(\ref{uni}) is not universal either. As the data available does
not seem enough to trace its correlations with other parameters, we
do not discuss the behavior of $\beta$ in the present paper.

\subsection{The high-field state of the material}   \label{hf}

The assumption of a model \cite{Fisher} that the high-field phase is
insulating was never questioned in experiments. It was always
presumed to be true because, according to the single-parameter
scaling theory \cite{4}, all 2D systems should be insulating at zero
temperature. Now this statement is questioned both experimentally
\cite{scal,scal1,Pud} and theoretically \cite{Dobr}. Subject to the
validity of the scaling hypothesis, there should exist a crossover
temperature $T^*$ from metallic to insulating behavior \cite{KhL},
i.e., from logarithmic quantum correction to the film conductance

\begin{equation}
\Delta\sigma\approx (e^2/\hbar)\ln(L(T)/l)<\sigma,\qquad\mbox{if}
\qquad T>T^* ,\label{met}
\end{equation}
where $L$ is the phase breaking length and $l$ is the elastic mean
free path, to activated conductance

\begin{equation}
\sigma\propto\exp[-(T_0/T)^p],\qquad\mbox{if}\qquad T<T^* ,
\label{ins}
\end{equation}
with parameters $T_0$ and $p=1$ or 1/2 or 1/3.

\begin{figure}[tb]
\vbox{
\epsfxsize=\figurewidth
\centerline{\epsfbox{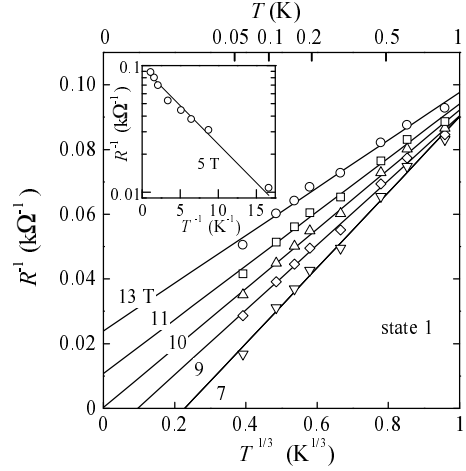}}
\caption{Temperature dependence of the high-field conduc\-tance of
state 1 at various magnetic fields. Arrhenius plot of the conductance
at $B=5$~T is displayed in the inset.}
\label{17}
}
\end{figure}

The inset to Fig.\,\ref{17} displays that, for state 1, the
resistance maximum in a field of 5~T does follow the activation
behavior (\ref{ins}) with $p=1$ and $T_0\approx 0.13\,$K. However, at
higher fields the activation law does not hold, nor does the
expression (\ref{met}) for the 2D case \cite{rem}. That is why we
examine the high-field resistance of our film in terms of 3D material
behavior in the vicinity of metal-insulator transition \cite{4}. In
this case the electron-electron interaction is dominant and the bulk
conductivity $\sigma$ should follow a power law \cite{Im,Alt1}

\begin{equation}		\label{t3}
\sigma (T)=a+bT^{1/3},
\end{equation}
where the factor $b>0$ and the sign of the parameter $a$
discriminates between a metal and an insulator at $T\rightarrow 0$
\cite{ImOv,VFGa}. If $a>0$, it yields zero-temperature conductivity
$\sigma(0)\equiv a$, whereas the negative $a$ points to activated
conductance at lower temperatures. From this standpoint, state 1,
which is superconducting below $B_c=2$\,T, is metallic above
$B=10$\,T and insulating in the intermediate field range
(Fig.\,\ref{17}). We believe that our material becomes metallic
because the field destroys localized Cooper pairs; this point
deserves a special discussion that will be given elsewhere.

\begin{figure}[tb]
\vbox{
\epsfxsize=\figurewidth
\centerline{\epsfbox{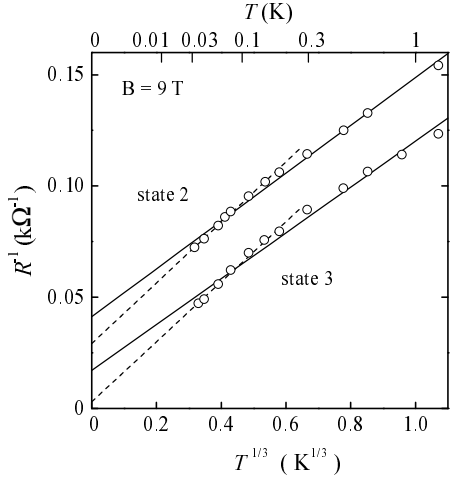}}
\caption{Temperature dependences of the conductance of states 2 and 3
in a magnetic field of 9~T. The high (low) $T$ data extrapolations
are shown by solid (dashed) lines.}
\label{18}
}
\end{figure}

The temperature dependence of the resistance for state 2 is very
different. Even at a field of 9~T, at the resistance maximum, the
value of $a$ is positive (Fig.\,\ref{18}). As seen from the figure,
with lowering $T$ the dependence $R(T)$ becomes stronger, resulting
in a lower but still positive $\sigma (0)$. Hence, we deal with the
transition from a superconductor to a normal metal. Finally, state 3
studied in the parallel field configuration is boundary for the two
above since the extrapolated value of $a$ is close to zero
(Fig.\,\ref{18}).

It is interesting that, despite on the high-field side of the
transition state 1 is insulating and state 2 is metallic, both of
these obey the scaling, see Figs.\,\ref{16}(b) and \ref{lin_sc}.

\subsection{2D or 3D ?}

The 3D behavior of the film resistance at high fields urges us to
dwell upon the question about the sample dimensionality. Apparently,
we have to compare the film thickness $d$ with characteristic
lengths. If far from the phase transition, these are the mean free
path $l$ in normal state and the coherence length $\xi_{sc}$ in
superconducting state. The former can be derived from the value of
conductivity $\sigma$ at $T\approx 4$\,K. In amorphous metals the
mean free path is normally close to the lowest possible value
$l\approx k_F^{-1}$. Knowing the normal state film resistance
$R\approx 5$\,k$\Omega$ and assuming that we deal with the metallic
3D material, we use the formulae

\begin{eqnarray}
\label{sig}
\sigma=&&ne^2l/\hbar k_F= \nonumber \\
&&(1/3\pi^2)(e^2/\hbar)(k_F l)^2/l\buildrel {l\rightarrow
1/k_F}\over\longrightarrow (1/3\pi^2)(e^2/\hbar)/l,\\
&&\qquad n=k_F^3/3\pi^2\approx(3\pi^2l^3)^{-1}
\end{eqnarray}
to estimate the electron density $n\approx 6\times 10^{19}\,{\rm
cm}^{-3}$ and the length $l\approx 8$~\AA. The value of $l$ is indeed
small compared to $d=200$~\AA.

The coherence length $\xi_{sc}$ in superconducting state can be
evaluated from the expression

\begin{equation}		\label{hc2}
\xi_{sc}=(\Phi_0/2\pi B_{c2})^{1/2},\qquad \Phi_0=\pi c\hbar/e.
\end{equation}
If we estimate the field $B_{c2}$ at $B_c=7.2\,$T as determined for
state 2, we get $\xi_{sc}\approx 70$\,\AA, which is appreciably
smaller than $d$. Hence, both estimates favour 3D.

However, both $l$ and $\xi_{sc}$ are not relevant in the vicinity of
quantum transition where $d$ should be compared with the diverging
correlation length $\xi\propto (B-B_c)^{-\nu}$. This implies that, at
zero temperature, sufficiently close to the transition a slab becomes
2D. At finite temperatures the spatial correlations linked with
quantum fluctuations are restricted by the dephasing length $L_\phi$
\cite{Girvin,QPT}

\begin{equation} \label{Lphi}
L_\phi\simeq\hbar^2/m\xi_{sc}T,
\end{equation}
where $\xi_{sc}\approx 100$~\AA\ stands for the short-distance
cutoff. That at $T=30$\,mK the value of $L_\phi\gg d$ is in favour of
the 2D model. This estimate is in contrast to the experimental
observation of the 3D behaviour near the metal-insulator transition
(Figs.\,\ref{17},\ref{18}), i.e., at our temperatures the phase
breaking length $L$ determined by electron-electron interactions is
small compared to $d$. We note that while the expression (\ref{t3})
fits the data for state 1 very well (Fig.\,\ref{17}), the data for
states 2 and 3 show the tendency to a stronger temperature dependence
at low $T$ (Fig.\,\ref{18}). This is likely to point to the fact that
for states 2 and 3 in the low-temperature limit the length $L$
approaches $d$, stimulating the localization.

In the end, if the QPT were indeed 2D, one would expect the
appreciable anisotropy of magnetoresistance with respect to the field
direction. We have checked that the two close states 2 and 3 of the
film studied, respectively, in normal and parallel fields behave
similarly (see Table~\ref{t1}). At least, there is no drastic effect
caused by changing the field direction and so the possibility of a 3D
field-tuned quantum SIT should be taken seriously.

\subsection{The role of disorder in QPT}

\begin{figure}[tb]
\vbox{
\epsfxsize=\figurewidth
\centerline{\epsfbox{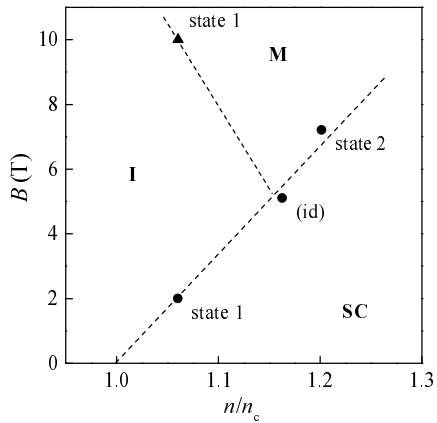}}
\caption{Schematic phase diagram of zero-temperature field-induced
transitions in the ($n,B$) plane. Super\-conduc\-ti\-vity destroying
transitions are marked by circles and an insulator-metal transition
from Fig.~\protect\ref{17} is marked by triangle. The dashed lines
are guides to the eye.}
\label{diagram}
}
\end{figure}

The phase diagram in Fig.\,\ref{diagram} summarizes the obtained data
on zero-temperature transitions. The zero-field SIT brought about by
changing the carrier density $n$ transforms above the critical
density $n_c$ into the field-induced SIT which turns into the
superconductor -- normal metal transition with further increasing
$n$.

Two significant factors allow us to distinguish the
magnetic-field-induced QPT at zero temperature \cite{QPT} from the
ordinary thermodynamic transition in superconductors at field
$B_{c2}(T)$: first, the symmetry with respect to the field $B_c$,
i.e., negative (positive) curvature of $R(T)$ below (above) $B_c$
(Fig.\,\ref{13}); second, the absence of features on the curves
$R(T)$ that can be identified as the onset of superconducting
transition. On the other hand, from the experimental data it follows
that the phenomenon of QPT is more general than the one considered in
Ref.~\cite{Fisher}. The superconducting phase can be alternated by
both an insulating phase (Fig.\,\ref{17}) and a metallic phase
(Fig.\,\ref{18}). The magnetic field direction and the reduced
dimension do not seem crucial. The finite slope of the separatrix
$R(T,B_c)$ implies that $R_c$ and/or $B_c$ depend on temperature.

So far, the quantum SIT was reliably observed only in amorphous films
\cite{HebPaa,Kapit,Gold}, although much work has been done on the
subject using different materials. Since in amorphous films a strong
disorder is expected, this gives a hint that in the limit of weak
disorder the magnetic-field-driven transition from a superconductor
to a normal metal should be thermodynamic with critical field
$B_{c2}(T)$ while the strong disorder is necessary for the QPT to
occur. Apparently, this issue demands further investigations.

\section{Conclusion}

In summary, we have studied the magnetic-field-tuned
superconductivity destroying QPT in amorphous In--O films with the
onset of superconductivity in zero field at $T_{c0}\approx 2$~K. To
obtain collapse of the data $R(T,B)$ in the vicinity of transition
against scaling variable $(B-B_c)/T^{1/y}$ we have either to take
account of the intrinsic temperature dependence of $R_c$ or to
postulate the temperature dependence of $B_c$. It has been found that
the magnetic field direction and the reduced dimension are not
crucial and that the state on the high-field side of the transition
can be both insulating and metallic. Presumably, it is the degree of
disorder that determines whether the field-induced superconductor --
normal metal transition is quantum or thermodynamic.

\acknowledgements

This work was supported by Grants RFBR~96-02-17497, RFBR~97-02-16829,
and INTAS-RFBR~95-302 and by the Programme "Statistical Physics" from
the Russian Ministry of Sciences.


\begin{references}
\bibitem{rem} This is in contrast to the fact that amorphous Mo--Ge
films from Ref.~\protect\cite{Kapit} demonstrated at high fields the
metallic temperature dependence in accordance with
Eq.~(\protect\ref{met}).
\end{references}
\end{document}